\documentclass[10pt,a4paper,twocolumn]{article}
\usepackage{ol2}
\usepackage[colorlinks=true,citecolor=blue]{hyperref}
\usepackage{amsmath}
\usepackage{lipsum}
\usepackage{graphics}
\usepackage{graphicx}
\begin{document}

\twocolumn[ %% activate for two-column option

\title{Spin orbit interaction of light mediated by scattering from plasmonic nano-structures}

%% For REVTeX it is possible to automate superscript and e-mail callouts with the superscriptaddress option; see REVTeX4 documentation.

\author{Jalpa Soni$^1$, Sayantan Ghosh$^{1}$, Shampy Mansha$^1$, S. Dutta Gupta$^{2,\$}$, Ayan Banerjee$^{1,\dagger}$ and Nirmalya Ghosh$^{1,*}$}

\address{
$^1$Dept. of Physical Sciences, IISER-Kolkata, BCKV Main Campus, Mohanpur 741 252, India
\\
$^2$School of Physical Sciences, University of Hyderabad, Hyderabad 500 046, India\\
Corresponding authors: $^\$$sdgsp@uohyd.ernet.in, $^\dagger$ayan@iiserkol.ac.in, $^*$nghosh@iiserkol.ac.in
}

\begin{abstract}The spin orbit interactions (SOI) of light mediated by single scattering from plasmon resonant metal nanoparticles (nanorods and nanospheres) are investigated using explicit theory based on Jones and Stokes-Mueller polarimetry formalism. The individual SOI effects are analyzed and interpreted via the Mueller matrix-derived, polarimetry characteristics, namely, diattenuation $d$ and retardance $\delta$. The results demonstrate that each of the contributing SOI effects can be controllably enhanced by exploiting the interference of two neighboring modes in plasmonic nanostructures (orthogonal electric dipolar modes in rods or electric dipolar and quadrupolar modes in spheres).\\
\end{abstract}


 ] %% activate for two-column option
Spin orbit interaction (SOI) of light has evoked intensive theoretical and experimental investigations in the past few years owing to fundamental interests and potential applications in nano-optics \cite{berry1987,hosten2008,bliokh2008,haefner2009,Dogariu:06,oscar2010,marrucci2006,zhao2007}. The SOI phenomenon signifies interconversion between the spin angular momentum (SAM, circular polarization represented by polarization helicity $\sigma=\pm1$) and orbital angular momentum (OAM, helical phase of light represented by topological vortex charge $l=0,\pm1,\pm2,\ldots$), which can be mediated via a number of processes involving light-matter interactions \cite{hosten2008,bliokh2008,haefner2009,Dogariu:06,oscar2010,marrucci2006,zhao2007}. Among these, while SOI arising due to propagation through anisotropic media is well known and studied over decades \cite{allen2003}, that occurring in isotropic media has only been noticed recently \cite{hosten2008,bliokh2008,haefner2009,Dogariu:06,oscar2010,marrucci2006,zhao2007}. For example, such intrinsic coupling between SAM and OAM has been observed in tight focusing of fundamental or higher order Gaussian beams \cite{zhao2007}, high numerical aperture (NA) imaging in isotropic homogenous/inhomogeneous media \cite{oscar2010} and in scattering from micro and nano systems \cite{haefner2009,Dogariu:06}. The spin orbit coupling in such cases have accordingly been modeled using different methods such as Debye Wolf theory for focusing \cite{oscar2010,Bomzon:07,Richards1959}, Mie theory for scattering \cite{haefner2009,Dogariu:06}. A number of interesting and intricate polarization effects associated with topological evolution of phase have also been observed and explained by SOI and subsequent conservation of total angular momentum (TAM) of light \cite{hosten2008,haefner2009,oscar2010,PhysRevB.82.125433,bliokh:2008,Schwartz:08}. Note each of the individual effects associated with SOI are manifested as a measurable change in the spatial polarization characteristics of light. For conceptual and practical reasons, modeling SOI via the conventional polarimetry characteristics may thus prove to be useful \cite{kliger1990}. In this letter, we present a generalized theory based on conventional Jones and Stokes-Mueller formalism for analysis, interpretation and quantification of the SOI effects (mediated by scattering) via the individual medium polarimetry characteristics, namely, diattenuation $d$ (differential attenuation of orthogonal polarization states) and retardance $\delta$ (phase shift between orthogonal polarization states)\cite{kliger1990,NGhosh:2011}. Importantly, we show using explicit theory and illustrative examples that spin orbit interactions by scattering can be significantly enhanced in plasmon resonant metal nanoparticles/nanostructures (nanorods and nanospheres). Moreover, in such nano-systems, each of the contributing SOI effects can be desirably tuned (optimized/enhanced) via the diattenuation and the retardance parameters by changing the wavelength of light and controlling the size and shape of the
nanoparticles.
\par
In order to study SOI of light mediated by the scattering process, let us choose the right handed Cartesian coordinate system with the incident light (plane wave) propagating in the $Z$ direction and assume that $X$ and $Y$ are the two orthogonal axes, representing the polarization axes in the laboratory reference frame. The scattered electric field $(\vec{E}^s)$ components can be related to the incident field $(\vec{E}^i)$ components in the laboratory frame by the action of the following transfer function $(J)$
\begin{eqnarray}
&\vec{E}^s\approx T_z(-\phi)T_y(-\theta)S(\theta)T_z(\phi)\vec{E}^i=J\vec{E}^i, \textrm{where,} \nonumber\\
& J=\begin{pmatrix}E_\alpha+E_\beta\cos2\phi&E_\beta\sin2\phi&E_\gamma\cos\phi\\
E_\beta\sin2\phi &E_\alpha-E_\beta\cos2\phi &E_\gamma\sin\phi\\
-E_\gamma\cos\phi&-E_\gamma\sin\phi &E_\alpha+E_\beta
\end{pmatrix}\label{eq:eq1}
\end{eqnarray}
where, $\theta$ is the scattering angle and $\phi$ is the azimuthal angle. The transformation matrix $T_z(\phi)$ transforms the incident laboratory frame field vector to the scattering plane. The inverse transformation matrices $T_z(-\phi)$ and $T_y(-\theta)$ transforms the scattered field from the scattering co-ordinate $(r,\theta,\phi)$ to the laboratory co-ordinate $(X,Y,Z)$. The matrix $S(\theta)$ (defined in the scattering plane) includes the effect of the scattering process in its elements; $S_2(\theta)$ - scattered field polarized parallel to the scattering plane$(\hat{\theta})$, and $S_1(\theta)$ - polarized
perpendicular to the scattering plane $(\hat{\phi})$\cite{Bohren:2007}. The scattered field descriptors, $E_\alpha(\theta)$, $E_\beta(\theta)$ and $E_\gamma(\theta)$ of Eq. \ref{eq:eq1}, are related to the elements of $S(\theta)$ as
\begin{equation}
E_\alpha=S_2\cos\theta+S_1;E_\beta=S_2\cos\theta-S_1;E_\gamma=-S_2\sin\theta
\label{eq:eq2}
\end{equation}
Note that the electric field transformation matrix of Eq. \ref{eq:eq1} is similar in nature to that for focusing \cite{oscar2010,zhao2007,Bomzon:07}. Thus akin to focusing, the SOI effect (SAM to OAM conversion) can readily be verified by applying the Jones vector $[1~\imath~0]^T$ of right circularly polarized (RCP) light on the matrix of Eq. \ref{eq:eq1} and by decomposing the resulting field into three uniform polarization components (as previously done for focusing of fundamental Gaussian laser beam \cite{Bomzon:07}). However, unlike focusing, the matrix $S(\theta)$ (in Eq. \ref{eq:eq1}) additionally incorporates information on the interactions of light with micro or nano scale objects and thus the SOI mediated by scattering from such systems is expected to be more complex. Since in an actual experiment, the set of analyzing optics and detectors (kept in the $X-Y$ plane, in far field) typically detect the transverse components of the scattered fields, henceforth we consider the transverse
components of the field only. The corresponding transfer function (first two rows and columns of Eq. \ref{eq:eq1}) is accordingly the conventional $2\times 2$ Jones matrix \cite{kliger1990}. In order to interpret the SOI via the conventional polarization parameters (diattenuation and retardance), we then proceed to derive the Mueller matrix corresponding to this Jones matrix, using standard relationship connecting Jones and Mueller-Jones matrices \cite{kliger1990,NGhosh:2011}. The resulting matrix is a diattenuating retarder Mueller matrix characterized by diattenuation $(d(\theta))$, retardance $(\delta(\theta))$ and orientation angle of the axes of the diattenuating retarder $\phi$ 
\begin{eqnarray}
&M(\theta,\phi)=\begin{pmatrix}
1&M_{12}&M_{13}&0\\
M_{12}&M_{22}&M_{23}&-M_{24}\\
M_{13}&M_{23}&M_{33}&M_{34}\\
0&M_{24}&-M_{34}&M_{44}\\
\end{pmatrix}
\label{eq:eq3}\\
&M_{12}=d\cos(2\phi),M_{13}=d\sin(2\phi)\nonumber\\
&M_{22}=\cos^2(2\phi)+x\cos(\delta)\sin^2(2\phi)\nonumber \\
&M_{23}=\sin(2\phi)\cos(2\phi)-x\cos(\delta)\sin(2\phi)\cos(2\phi)\nonumber\\
&M_{24}=x\sin(\delta)\sin(2\phi)\nonumber\\
&M_{33}=\sin^2(2\phi)+x\cos(\delta)\cos^2(2\phi)\nonumber\\
&M_{34}=x\sin(\delta)\cos(2\phi),M_{44}=x\cos(\delta),x=|\sqrt{1-d^2}|\nonumber
\end{eqnarray}
Here, the $d$ and $\delta$ parameters are related to $E_\alpha(\theta)$, $E_\beta(\theta)$, and thus to the amplitude scattering matrix elements, $S_2(\theta)$ and $S_1(\theta)$ as
\begin{eqnarray}
d(\theta)=&\left\{\frac{\vert S_2(\theta)\vert^2\cos^2\theta-\vert S_1(\theta)\vert^2}{\vert S_2(\theta)\vert^2\cos^2\theta+\vert S_1(\theta)\vert^2}\right\},\nonumber\\\delta(\theta)=&\sin^{-1}\left[\frac{2\textrm{Im}\left(S_2^*(\theta)S_1(\theta)\right)}{\vert S_2(\theta)\vert\vert S_1(\theta)\vert}\right]
\label{eq:eq4}
\end{eqnarray}
In what follows, we analyze and interpret the competing three individual SOI effects via the pure Mueller matrices corresponding to each of these effects using the d and δ parameters encoded in its various elements.
\section*{\emph{Case 1:} $d=0, \delta=\pi \leftrightarrow E_\alpha=0$ or $S_2\cos\theta=-S_1$;Geometrical Berry phase effect}
The Mueller matrix (Eq. \ref{eq:eq3}) corresponding to this special case represents pure geometrical Berry phase effect associated with azimuthal $(\phi)$ rotation of polarization (spin) and subsequent generation of geometrical phase vortex (OAM) \cite{berry1987,marrucci2006,Bomzon:07}. The nature of SOI for such topological phase evolution can be understood by applying the Stokes vector corresponding to a horizontal linear polarization state $(\mathbf{S}_i=[1~1~0~0]^T)$ on this matrix. The output Stokes vector becomes: $\mathbf{S}_o = M\mathbf{S}_i = [1~\cos 4\phi~\sin 4\phi~0]^T$ ,which implies rotation of the incident linear polarization state by twice the azimuth
angle $(2\phi)$. This can be interpreted as, the two circular polarization modes (left and right), constituting the initial linear polarization state, acquire opposite phase vortices $(l=\pm 2)$ during their evolution in the scattering process \cite{Bomzon:07,Schwartz:08}. Any incident circular polarization state (e.g, LCP / RCP), on the other hand, undergoes flipping of helicity $\sigma=+1\leftrightarrow-1$ (evident from the negative sign of the matrix element $M_{44}$ under this condition), and subsequently generates phase vortex with topological charge $l=\pm 2$. This phenomenon has been observed in backscattering from random medium \cite{Schwartz:08}. Note, however, the condition $S_2\cos\theta=-S_1$ can never be fulfilled for single scattering from dielectric Rayleigh scatterers  (radius $r<<\lambda$, where scattering is primarily contributed by the lowest order TM scattering $a_1$ mode \cite{Bohren:2007})
either for forward scattering $(0 \leq \theta \leq \pi/2)$ or for backscattering $(\pi/2 \leq \theta \leq \pi)$ angles. In contrast, this condition can indeed be satisfied
even for small angle (forward) scattering from plasmon resonant metal nanorods\cite{maier2007}. In fact, all the competing SOI effects (discussed subsequently) can be significantly enhanced in controllable fashion, as illustrated below with selected examples.
\par
\begin{figure}[h]
\centering
\includegraphics[width=0.85\columnwidth]{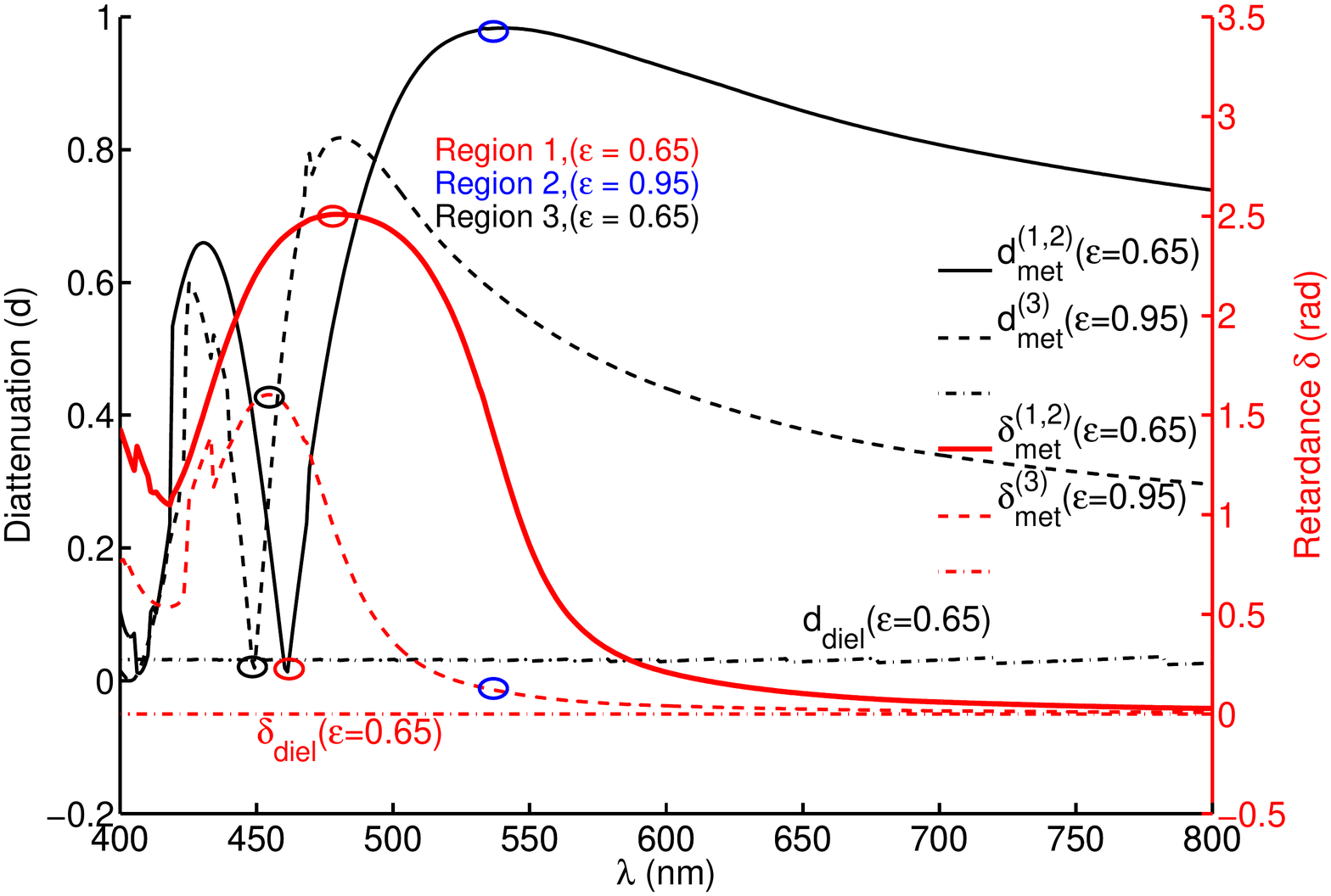}
\caption{\label{fig:fig1}(color online) Variations of diattenuation $d$ (left axis, black lines) and retardance $\delta$ (right axis, red lines) with wavelength $\lambda$ for silver nanorods having equal volume sphere radius $r=20nm$ and for two aspect ratio values $\varepsilon=0.65$ (corresponding $d$ and $\delta$: solid lines) and 0.95 ($d$ and $\delta$: dashed lines). The three different regions satisfying conditions for Case - 1, 2 and 3 SOI effects are marked as Region 1 (red), 2 (blue) and 3 (black) respectively. Conditions 1 and 2 are satisfied at two different wavelengths for metal nanorods with $\varepsilon=0.65$ ($d_{\textrm{met}}^{(1,2)}$ and $\delta_{\textrm{met}}^{(1,2)}$) and condition 3 is satisfied for metal nanorods with $\varepsilon=0.95$ ($d_{\textrm{met}}^{(3)}$ and $\delta_{\textrm{met}}^{(3)}$). The magnitudes of $d$ (black dotted line) and $\delta$ (red dotted line) for similar dielectric nanorods ($r=20nm, \varepsilon=0.65$) are considerably weaker and do not exhibit spectral dependence.}
\end{figure}

Briefly, metal nanorods exhibit two electric dipolar plasmon
resonances, one at shorter wavelength (transverse resonance along the short axis) and the other at longer wavelength (longitudinal resonance along the long axis) \cite{maier2007}. In Figure \ref{fig:fig1}, we show the wavelength variation $(λ = 400nm- 800nm)$ of the computed $d$, and $\delta$ parameters for preferentially oriented silver nanorods having equal volume sphere radius $r = 20 nm$ and for two different aspect
ratios of the rods (ratio of diameter to length) $\varepsilon = 0.65$ and $0.95$ respectively. The scattering angle is chosen to be $\theta=15^\circ$ (as representative forward scattering small angle). The nanorods were oriented such that the long and the short axes of the rods are aligned along the laboratory $X-Y$ (polarization) axes respectively\cite{Mishchenko:98}. The scattering matrices of the rods were calculated using T-matrix \cite{Mishchenko:98} method and the $d$ and $\delta$ parameters were subsequently determined employing Eq. \ref{eq:eq4}.
\par
Several observations are at place. The diattenuation $d$ parameter peaks at wavelengths corresponding to the two electric dipolar plasmon resonance bands (transverse at $\lambda \sim 425 nm$, longitudinal at $\lambda \geq 500 nm$)\cite{maier2007}. In contrast, at the spectral overlap region of the two
resonances $(\lambda \sim 450 - 500 nm)$, $d$ exhibits sharp concavity $(d \sim 0)$. This rapid variation of diattenuation is a manifestation of preferential excitation of the two resonances with respective orthogonal linear polarizations for such preferentially oriented rods \cite{jsoni2012}. In addition to diattenuation, the nanorods also exhibit strong linear retardance $\delta$ effect, the magnitude of which peaks around the spectral overlap region of the two resonances. The enhanced $\delta$ effect can be attributed to the inherent phase retardation between the two competing electric dipolar plasmon modes \cite{jsoni2012}. Importantly, the magnitudes of $d$ and $\delta$ can be desirably controlled by changing the aspect ratio $\varepsilon$ of the rods. In fact, at the overlap region of the two resonances $(\lambda \sim 480 nm)$, the parameter $\delta$ and $d$ can be tuned (by reducing $\varepsilon$ to $\sim0.65$) so that they satisfy the condition $\delta \sim \pi$ , $d \sim 0$ (implying pure helicity reversal in forward scattering) (marked in red color, as region 1 in Figure \ref{fig:fig1}). Clearly, such effect can never be achieved from dielectric Rayleigh scatterers (shown in Figure \ref{fig:fig1}), and is exclusively enabled by the presence of the two neighboring dipolar plasmon resonance modes in the metal nanorods.
\section*{\emph{Case 2:} $d=1, \delta\sim 0, \leftrightarrow E_\alpha=\pm E_\beta$ or $S_2=0/S_1=0$; SAM to OAM conversion via pure diattenuation effect of scattering}
The Mueller matrix (Eq. \ref{eq:eq3}) corresponding to this special case represents pure diattenuator matrix and signifies complete conversion of SAM to OAM $(\sigma = \pm1 \rightarrow 0, l = \pm 1)$. This can be verified by applying the Stokes vector corresponding to input RCP state $(\mathbf{S}_i=[1~ 0~ 0~ 1]^T, \sigma = +1)$ on this matrix. The output Stokes vector represents pure linear polarization state $(\sigma = 0): \mathbf{S}_o = [1~ cos(2\phi)~ sin (2\phi)~ 0]^T$. The generation of $l = +1$ vortex is manifested in the second and the third (linear polarization descriptor) elements of the output Stokes vector, by the appearance of the $\cos2\phi$ and $\sin2\phi$ factors. The scattered light thus becomes completely linearly polarized, carrying no SAM and accordingly the angular momentum is entirely carried by the OAM term $(\sigma = +1\rightarrow \sigma = 0, l = +1)$. As shown in Figure \ref{fig:fig1}, the condition for such SOI $(d=1,\delta \sim 0)$ can be satisfied for plasmonic nanorods at wavelengths corresponding to the peaks of the dipolar plasmon resonances (at $\lambda \sim 525 nm$ corresponding to the longitudinal dipolar resonance shown here, in blue color, as region 2). Note that such effect can arise even for scattering from dielectric Rayleigh scatterers\cite{Dogariu:06}. The condition is fulfilled for Rayleigh scatterer at scattering angle $\theta=90^\circ$, where the amplitude scattering matrix element $S_2$ vanishes, yielding $d = 1, \delta = 0$. For large sized dielectric Mie scatterers $(r \geq \lambda)$ such complete conversion of SAM to OAM may occur at several other narrow range of angles depending upon the size parameter of scatterer \cite{haefner2009,Dogariu:06}. 
\begin{figure}
\centering
\includegraphics[width=0.85\columnwidth]{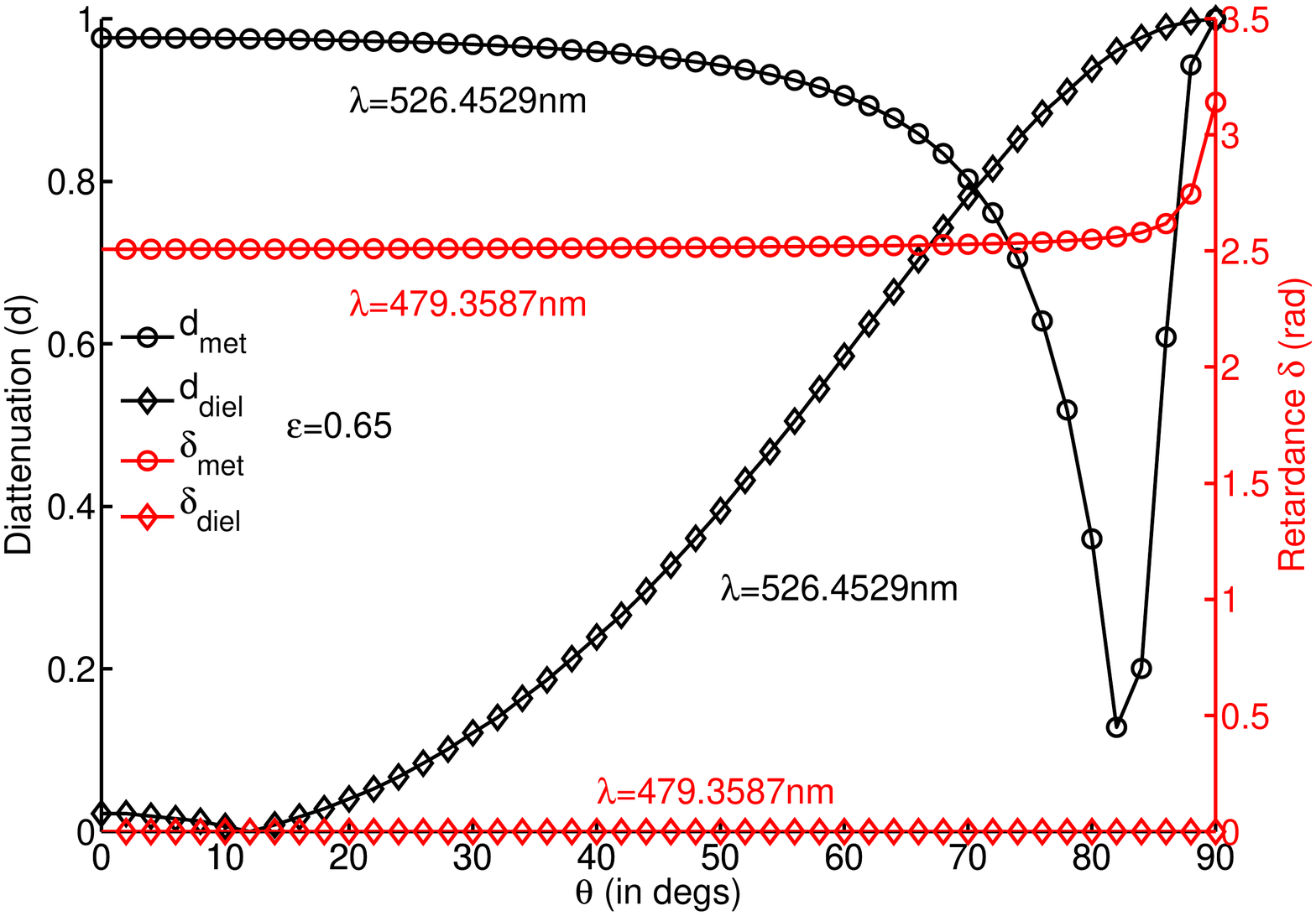}
\caption{\label{fig:fig2}(color online) Variations of diattenuation $d$ (left axis, black circle) and retardance $\delta$ (right axis, red circle) as a function of scattering angle $\theta$ for silver nanorods $(r = 20 nm, \varepsilon = 0.65)$. The magnitude of $d$ at $\lambda = 526 nm$ (peak of longitudinal plasmon resonance) is $\sim 1$ for a broad range of $\theta$ (satisfies condition for Case 2 SOI). In contrast, $d$ for similar dielectric nanorods $(r = 20 nm, \varepsilon = 0.65)$ (black triangle) is quite low at smaller $\theta$ and approaches unity only at $\theta=90^\circ$. The magnitude of $\delta$ for silver nanorod at $\lambda = 479 nm$ (overlap region of the two dipolar plasmon resonances) remain close to $\pi$ over a broad range of $\theta$ (satisfies condition for Case 1 SOI).}
\end{figure}
In contrast, for the plasmonic nanorods, the phenomenon can be observed over a broad range of forward scattering angles, as illustrated in Figure \ref{fig:fig2}.
\par
The angular variation of the derived $d$ and $\delta$ parameters for the silver nanorod with $r=20 nm, \varepsilon=0.65$, are shown here in Figure \ref{fig:fig2}, for two different wavelengths, $\lambda~525 nm$ (longitudinal resonance peak) and $\lambda~480 nm$ (overlap region of two dipolar modes). Evidently, the condition $d=1, \delta=0 (\text{for} \lambda~525 nm)$ is satisfied over almost the entire range of forward scattering angles.
\section*{\emph{Case 3:} $d=0,\delta=\pi/2,\leftrightarrow E_\alpha=\imath E_\beta$ or $S_2\cos\theta=\imath S_1$; SAM to OAM conversion via pure retardance effect of scattering}
In this case, the resulting Mueller matrix assumes the form of a pure retarder matrix. For input horizontal linear polarization state $(\mathbf{S}_i=[1~ 1~ 0~ 0]^T)$, the output Stokes vector of the scattered light becomes $\mathbf{S}_o= [1~ 1~ \frac{1}{2}(1 + \cos 4 \phi )~ \frac{1}{2}\sin 4 \phi~ \sin 2 \phi ]^T$ , implying generation of azimuthal angle $\phi$ - separated lobes of opposite circular polarization states. This effect is similar in nature to the Spin Hall effect of light, wherein an incident linear polarization state, evolve in different trajectories to generate spatially separated lobes of opposite circular polarization states $(\sigma = \pm 1)$\cite{oscar2010}. Note in this case, the SAM to OAM conversion for input circular polarization state (e.g., RCP, $\mathbf{S}_i = [1~ 0~ 0~ 1]^T$ ), is similar to that observed for the previous effect. The output Stokes vector $\mathbf{S}_o = [1 ~- \sin (2\phi) ~\cos(2\phi)~ 0]^T$ implies complete SAM to OAM conversion and subsequent generation of phase vortex $l = +1 (\sigma = +1\rightarrow \sigma = 0, l = +1)$. From Figure \ref{fig:fig1}, it is apparent that the condition for this type of SOI can be fulfilled in the spectral overlap region $(\lambda \sim 480 nm)$ of the two dipolar plasmon resonances, for relatively larger aspect ratio of the silver nanorod $(\varepsilon= 0.95)$ (shown in blue color, as region 2 in figure \ref{fig:fig1}).
\begin{figure}
\centering
\includegraphics[width=0.85\columnwidth]{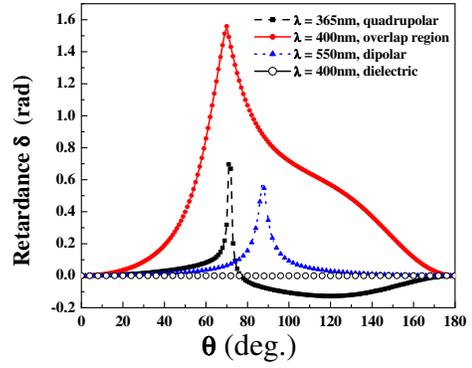}
\caption{\label{fig:fig3}(color online) Variations of retardance δ as a function of scattering angle $\theta$ for silver nanosphere $(r = 50 nm)$, for three different wavelengths, $\lambda=365nm$ (quadrupolar resonance peak, black dashed line), $550 nm$ (dipolar resonance peak, blue dotted line) and $400 nm$ (overlap spectral region of the
two modes, red solid line). Significant enhancement of magnitude of $\delta$ is apparent at $\lambda = 400 nm$. The $\delta$-value at $\lambda = 400 nm$ for a dielectric sphere having identical size (shown by circle) is rather low.}
\end{figure}
\par
The results presented above show that enhancement of the SOI
effects can be achieved by simultaneously exciting the two electric dipolar plasmon modes of metal nanorods. In general, such resonant enhancement of SOI should be possible in metal nanospheres also by exiting neighboring dipolar and quadrupolar plasmon modes. As an illustrative example, in Figure \ref{fig:fig3}, we show enhancement of the retardance $\delta$ parameter by simultaneous excitation of the electric dipolar $(a_1)$ and electric quadrupolar $(a_2)$ plasmon modes for a silver sphere with radius $r = 50 nm$\cite{Bohren:2007,maier2007}. For this metal sphere, the $a_1$ and the $a_2$ plasmon modes are characterized by resonance peaks at $550 nm$ and $365 nm$ respectively. The results are thus shown for three different wavelengths $(\lambda= 550 nm, 365 nm$, peaks of the dipolar and the quadrupolar resonances respectively, and $400 nm$, overlap region of the two modes). Apparently, $\delta$ shows significant enhancement at $\lambda= 400 nm$, where both the $a_1$ and $a_2$ plasmon modes are excited simultaneously. In contrast, a dielectric sphere having identical size (shown in the figure \ref{fig:fig3}) does not exhibit any appreciable value of this parameter, implying the corresponding SOI effect is rather weak in absence of the plasmon resonances.
\par
To summarize, the results demonstrate that each of the scattering mediated SOI effects can be resonantly enhanced by exploiting the interference of two neighboring plasmon modes in metal nanostructures (orthogonal electric dipolar modes in rods or electric dipolar and quadrupolar modes in spheres). Importantly, the contributing SOI effects can be desirably tuned (optimized/enhanced) by changing the wavelength of light and controlling the size, shape of the nanoparticles. The developed generalized framework based on Mueller matrix approach enabled quantification and interpretation of each of the individual SOI effects, via their characteristic signature in the polarization patterns
of the Mueller matrix elements (two-fold or four-fold azimuthal patterns depending upon the nature of the SOI), and via the diattenuation $d$ and retardance $\delta$ parameters encoded in the matrix elements. Although the present formalism is derived for scattering of plane waves, extension of this to include scattering of fundamental and higher order Gaussian beams, is also warranted. Since, the d and $\delta$ parameters can be directly determined from any experimental Mueller matrix \cite{NGhosh:2011}, these parameters hold promise as novel experimental metrics for studying spin orbit interactions mediated by interactions of light with micro or nano-scale objects. 
\bibliography{OLbib.bib}
\bibliographystyle{ol}
\end{document}